\documentclass[a4paper,11pt]{article}
\usepackage{pos}
\usepackage{bbm}
\usepackage{marvosym}
\usepackage{diagbox}

\title{Hamiltonian Lattice QCD from Strong Coupling Expansion}

\author[a]{Pratitee Pattanaik}
\author*[a]{Wolfgang Unger}

\affiliation[a]{Fakultät für Physik, Universität Bielefeld,\\
  Bielefeld, Germany}

\emailAdd{pratiteep@physik.uni-bielefeld.de}
\emailAdd{wunger@physik.uni-bielefeld.de}

\abstract{We present generalizations of Hamiltonian Lattice QCD as derived from the continuous time limit of strong coupling lattice QCD: we discuss the flavor dependence and the effect of gauge corrections. This formalism can be applied at finite temperature and baryon density as well as isospin density and allows both for analytic and numeric investigations that are sign problem-free.
}

\FullConference{%
 The 38th International Symposium on Lattice Field Theory, LATTICE2021
  26th-30th July, 2021
  Zoom/Gather@Massachusetts Institute of Technology
}

\def \Tr {{\rm Tr}}
\def \tr {{\rm tr}}

\newcommand{\expval}[1]{{\langle #1\rangle }}

\newcommand{\bareT}{\mathcal{T}}
\newcommand{\bareMu}{\mu_{\mathcal{B}}}

\newcommand{\meson}{\mathfrak{m}}
\newcommand{\baryon}{\mathfrak{b}}
\newcommand{\hadron}{\mathfrak{h}}
\newcommand{\spin}{\mathfrak{s}}

\newcommand{\SU}{{\rm SU}}

\def \at {{a_{\tau}}}

\def \Nt {{N_{\tau}}}
\def \Nc {{N_{c}}}
\def \Nf {{N_{f}}}

\newcommand{\Hil}{\mathbbm{H}_\hadron}

\newcommand{\calH}{\mathcal{H}}
\newcommand{\calN}{\mathcal{N}}

\newcommand{\lsteel}{{\hspace{0.1em}\text{\Lsteel}\hspace{0.1em}}}
\newcommand{\tsteel}{{\hspace{0.1em}\text{\Tsteel}\hspace{0.1em}}}

\newcommand{\nn} {\nonumber\\}

\begin{document}
\maketitle

\section{Introduction}

The Hamiltonian formulation of lattice QCD has been discussed in detail in \cite{Klegrewe2020} in the strong coupling limit for $\Nf=1$. In contrast to Hamiltonian formulations in the early days of lattice QCD \cite{Kogut1974} this formulation is based on the dual representation that is obtained when integrating out the gauge links first, and then the Grassmann-valued fermions \cite{Rossi1984}.
The resulting dual degrees of freedom are color singlets, such as mesons and baryons. In this representation, the finite density sign problem is much milder, as it is re-expressed in terms of the geometry of baryonic world-lines.
The dual representation of lattice QCD with staggered fermions has been studied in the strong coupling limit both via mean-field theory \cite{Kawamoto1981} and Monte Carlo \cite{Karsch1989,Forcrand2010} and has been extended in both approaches to include gauge corrections \cite{deForcrand2014,Gagliardi2019}. 

The Hamiltonian formulation is based on the continuous time limit of the dual representation: at fixed bare temperature $aT=\frac{a}{a_t N_t}$, the limits $a_t \rightarrow 0$ and $N_t\rightarrow \infty$ are taken simultaneously \cite{deForcrand2017}.
This limit requires to determine the non-perturbative relation between the bare anisotropy $\gamma$ and the physical anisotropy $\frac{a}{a_t}\equiv\xi(\gamma)$, based on pion current fluctuations.
A conservation law for the pion current can be identified: if a quantum number $\meson(x)$ is raised/lowered by a spatial dimer, then at the site connected by the spatial dimer the quantum number is lowered/raised. This is a direct consequence of the even/odd decomposition for staggered fermions. With interactions derived from a high temperature series, the resulting partition sum can be expressed in terms of a Hamiltonian that is composed of mesonic annihilation and creation operators $\hat{J}^\pm$:
\begin{align}
&Z_{\rm CT}(\bareT,\bareMu)=\Tr_\hadron\left[e^{(\hat{\calH}+\hat{\calN}\bareMu)/\bareT}\right],\qquad
\hat{\calH}=\frac{1}{2}\sum_{
\langle\vec{x},\vec{y}\rangle}
\left(
\hat{J}^{+}_{\vec{x}} \hat{J}^{-}_{\vec{y}}+
\hat{J}^{-}_{\vec{x}} \hat{J}^{+}_{\vec{y}}
\right),\qquad
\hat{\calN}=\sum_{\vec{x}}\hat{\omega}_x,\nn
&\hat{J}^+=
{\footnotesize
\left(
\begin{array}{cccc|cc}
0   & 0   & 0   & 0 &  & \\
\hat{v}_\lsteel & 0   & 0   & 0 &  & \\
0   & \hat{v}_\tsteel & 0   & 0 &  & \\
0   & 0   & \hat{v}_\lsteel & 0 &  & \\
\hline
 &  &  &  & 0 & 0\\
 &  &  &  & 0 & 0\\
\end{array}
\right)}
,\qquad
\hat{J}^{-}=(\hat{J}^{+})^T, \qquad
\hat{\omega}=
{\footnotesize
\left(
\begin{array}{cccc|cc}
0 & 0 & 0 & 0 &  & \\
0 & 0 & 0 & 0 &  & \\
0 & 0 & 0 & 0 &  & \\
0 & 0 & 0 & 0 &  & \\
\hline
 &  &  &  & 1 & 0\\
 &  &  &  & 0 & -1\\
\end{array}
\right)
}
\label{ParFuncHam}
\end{align}
with the local Hilbert space given by 
$| \hadron\rangle=| \meson;\baryon\rangle =|
0, \pi, 2\pi, 3\pi; B^+, B^- \rangle.$
The matrix elements $\hat{v}_\lsteel=1$ and $\hat{v}_\tsteel=\frac{2}{\sqrt{3}}$ are computed from the local weights of the corresponding meson vertices.
Since Pauli saturation holds on the level of the quarks and mesons have a fermionic substructure, the meson occupation numbers $|\meson\rangle$ are also bounded from above. This results in a particle-hole symmetry, leading to an SU(2) algebra which is $d=\Nc+1$-dimensional:
\begin{align*}
 \hat{J}_1&= \frac{\sqrt{\Nc}}{2}\left(\hat{J}^+ + \hat{J}^-\right), \quad \hat{J}_2= \frac{\sqrt{\Nc}}{2i}\left(\hat{J}^+ - \hat{J}^-\right), &
 \hat{J}_3&=i[J_1,J_2]
=\frac{\Nc}{2}[\hat{J}^+,\hat{J}^-],\quad
\hat{J}^2 = 
\frac{\Nc(\Nc+2)}{4}
,
\end{align*}
\vspace{-9mm}
\begin{align}
 \hat{J}_3 \left|\frac{\Nc}{2},\spin \right\rangle&=\spin \left|\frac{\Nc}{2},\spin \right\rangle,& \hat{J}^2 \left|\frac{\Nc}{2}\spin\right\rangle&=\frac{\Nc\left(\Nc+2\right)}{4} \left|\frac{\Nc}{2},\spin \right\rangle,& [\hat{J}^2,\hat{J}_3]&=0
\end{align}
with $\meson  \mapsto\spin=\meson-\frac{\Nc}{2}$.
The $\Nf=1$ formulation at strong coupling has many similarities with full QCD, and via Quantum Monte Carlo the grand-canonical and canonical phase diagram could be determined \cite{Klegrewe2020}. The nuclear interactions are of entropic nature: the nucleons, which are point-like in the strong coupling limit, attract each other due to the modification of the pion bath that surrounds them. 
However, pion exchange cannot be realized in the strong coupling limit for $\Nf=1$ since the Grassmann constraint does not allow for pions and nucleons to overlap. This is clearly a lattice artifact of the strong coupling limit, but this is overcome by formulations with $\Nf>1$ and/or by including gauge corrections.

The aim of the generalizations discussed in this proceedings is to allow for Quantum Monte Carlo simulations (QMC) at both non-zero baryon and isospin chemical potential, and possibly also strangeness chemical potential. 
We will argue that in the strong coupling limit, the $\Nf=2$ formulation remains sign problem-free. 

\section{Hamiltonian formulation in the strong coupling limit for any $N_f$}

\def \QMat{\mathcal{M}}
\newcommand{\I}{\mathcal{I}}
\newcommand{\J}{\mathcal{J}}
\newcommand{\Qmd}{\QMat^\dagger}
\newcommand{\M}{\mathcal{M}}
\newcommand{\Md}{\mathcal{M}^\dagger}   
\newcommand{\Ud}{U^\dagger}
\newcommand{\MMd}{\mathcal{M}\mathcal{M}^\dagger}
\newcommand{\Mxy}{M_{xy}}
\newcommand{\mmd}{\QMat\Qmd}
\newcommand{\floor}[1]{\lfloor #1 \rfloor}
\newcommand{\ceil}[1]{\lceil #1 \rceil}
\setlength{\tabcolsep}{2pt}

\newcommand{\kU}{k_U}
\newcommand{\kD}{k_D}
\newcommand{\kP}{k_{\pi^+}}
\newcommand{\kM}{k_{\pi^-}}
\newcommand{\kPPUD}{k^{(2)}_{\pi^+\pi^-,UD}}
\newcommand{\kUDPP}{k^{(2)}_{UD,\pi^+\pi^-}}
\newcommand{\mZ}{\mathfrak{m}_0}
\newcommand{\mPZ}{\mathfrak{m}_{\pi^0}^2}
\newcommand{\mSZ}{\mathfrak{m}_{\bar{\pi}^0}^2}
\newcommand{\mU}{\pi_U}
\newcommand{\mD}{\pi_D}
\newcommand{\mS}{\pi_S}
\newcommand{\mP}{\pi_+}
\newcommand{\mM}{\pi_-}
\newcommand{\KP}{K_+}
\newcommand{\KM}{K_-}
\newcommand{\mPPUD}{\mathfrak{m}_{\pi^+\pi^-,UD}}
\newcommand{\mUDPP}{\mathfrak{m}_{UD,\pi^+\pi^-}}
\renewcommand{\mZ}{\mathfrak{m}_0}
\renewcommand{\mPZ}{\pi_0^2}
\renewcommand{\mSZ}{\bar\pi_0^2}
\newcommand{\xS}{XS}

\newcommand{\lr}[1]{\left(#1\right)}
\newcommand{\blue}[1]{{\color{blue}{#1}}}
\newcommand{\red}[1]{{\color{red}{#1}}}

Strong coupling lattice QCD with staggered fermions for $\Nf>1$ admits more than one baryon per site and the Grassmann constraint allows for pion exchange between them, modifying nuclear interactions substantially. It has so far only been studied via mean-field theory \cite{Bilic1992a}. A Hamiltonian formulation for $\Nf=2$ allows for QMC and provides a more realistic scenario for nuclear interactions and the phase diagram.
It also compares better to the strong coupling regime with Wilson fermions in a world-line formulation, as discussed in the context of the 3-dim.~Polyakov effective theory (for a review, see \cite{Philipsen2021}).
Also for $\Nf>1$ the suppression of spatial bonds $\gamma^{-k}$, $k>2$ applies, hence the continuous time limit is well defined. 

The first step to derive the Hamiltonian is to determine the local Hilbert space $\Hil$ via canonical sectors $B \in \{-\Nf,\ldots, \Nf\}$, i.e.~we need to consider all possible single-site quantum states $\hadron$ in a non-interacting theory to establish the basis of quantum states that generalize the $\Nf=1$ states $|\meson\rangle$ and  $|\baryon\rangle$. 
The static partition sum for any $\Nc$, $\Nf$ is:
\begin{align}
Z_{\rm stat}(\Nc,\Nf)\sum_{B=-\Nf}^{\Nf}
    \prod_{a=0}^\Nc
    \frac{a!(2\Nf+a)!}{(\Nf+a+B)!(Nf+a-B)!}e^{B \mu_B/T}
\end{align}
where the coefficient encode the number of hadronic states, i.e.~the dimension of $\hadron$. For $\Nc=3$ (separated in canonical sectors)
\begin{itemize}
 \item for $\Nf=1$: $d=[1,4,1]=6$,
 \item for $\Nf=2$: $d=[1,20,50,20,1]=92$,
 \item for $\Nf=3$: $d=[1,56,490,980,490,56,1]=2074$. 
\end{itemize}
Unfortunately there is no general formula known including the isospin sectors, required for $\mu_I\neq 0$. The individual states $\hadron$ are also required for constructing the flavored $\hat{J}^\pm$.
To arrive at this basis, we consider the $\SU(3)$ one-link integrals in terms of the fermion matrix $(\QMat)_{ij}=\bar{\chi}^\alpha_{i}(x) \chi^\alpha_i(y)$ and $(\QMat^\dagger)_{kl}=\chi^\beta_k(y)\bar{\chi}^\beta_l(x)$, valid also for $\Nf\geq1$:
\begin{align}
 \J(\QMat,\Qmd) &=\int\limits_{\SU(3)}\hspace{-2mm} dU e^{\tr[ U \Qmd+U^\dagger \QMat]}= \sum_{B=-\Nf}^\Nf \sum_{n_1,n_2,n_3}
C_{B,n_1,n_2,n_3}
  \frac{E^B}{|B|!}\prod_{i=1}^3\frac{X_i^{n_i}}{n_i!},\quad E=
\begin{cases}
\det{\QMat}& B>0\\
1& B=0\\
\det{\Qmd}& B<0\\
\end{cases}
\nn
C_{B,n_1,n_2,n_3}&=2\binom{n_1+2n_2+4n_3+2|B|+2}{n_3+|B|}
\frac{|B|!}{(n_1+2n_2+3n_3+2|B|+2)!(n_2+2n_3+|B|+1)!}
\label{JInt}
\end{align}
which in contrasts to \cite{Eriksson1981} is now expressed in a more suitable basis involving the baryon sectors $B$.
The sum over $B$ and $n_i$ $(i=1,\ldots 3)$ terminates due to the Grassmann integration, depending on $\Nf$.
The corresponding invariants $D$, $X_i$ can be evaluated as follows:
\begin{align}
 X_1&=\tr[\mmd]=\Tr[M_x M_y]=\kU+\kD+\ldots + \kP+\kM + \ldots\nn
 X_2&=\frac{1}{2}\left(\tr[\mmd]-\tr[(\mmd)^2]\right)=\frac{1}{2}\lr{\Tr[M_x M_y]^2+\Tr[(M_x M_y)^2]}
 = X_1^2-D_2\nn
 X_3&=\det[\mmd]=\frac{1}{6}\lr{\Tr[M_x M_y]^3+3\Tr[M_x M_y] \Tr[(M_x M_y)^2] +2\Tr[(M_x M_y)^3]}
 =X_1^3-2 X_1 D_2+D_3\nn
D_2&= \frac{1}{2}\lr{\Tr[M_x M_y]^2-\Tr[(M_x M_y)^2]}=\kU\kD + \kP\kM +\ldots  - (\kPPUD+\kUDPP+\ldots)\nn
D_3&=\frac{1}{6}\lr{\Tr[M_x M_y]^3-3\Tr[M_x M_y] \Tr[(M_x M_y)^2] +2\Tr[(M_x M_y)^3]}=\frac{1}{6}(3X_1 D_2-X_1^3)\nn
E&=\det[\QMat]=\sum_{f\leq g\leq  h }B_{fgh}\quad (B>0)\quad  \text{or} \quad E=\det[\QMat^\dagger]=\sum_{f\leq g\leq  h }\bar{B}_{fgh}, \quad (B<0)
 \end{align}
 where the invariants are expressed in terms of nearest neighbors $(M_x M_y)^n = (-1)^{n+1} (\QMat\Qmd)^n$.
To be explicit, for $\Nf=2$, the (anti-) baryons and flavored dimers in terms of quarks are:
\begin{align*}
B_{uud}&=\bar{u}\bar{u}\bar{d}_x uud_y,&
B_{udd}&=\bar{u}\bar{d}\bar{d}_x udd_y,&
B_{uuu}&=\bar{u}\bar{u}\bar{u}_x uuu_y,&
B_{ddd}&=\bar{d}\bar{d}\bar{d}_x ddd_y,\nn
\bar{B}_{uud}&=uud_x\bar{u}\bar{u}\bar{d}_y,&
\bar{B}_{udd}&=udd_x\bar{u}\bar{d}\bar{d}_y,&
\bar{B}_{uuu}&=uuu_x\bar{u}\bar{u}\bar{u}_y,&
\bar{B}_{ddd}&=ddd_x\bar{d}\bar{d}\bar{d}_y,\nn
\kU&=\bar{u}u(x)\bar{u}u(y),&  \kD&=\bar{d}d(x)\bar{d}d(y),& \kP&=\bar{u}d(x)\bar{d}u(y), & \kM&=\bar{d}u(x)\bar{u}d(y)
\end{align*}
\vspace{-9mm}
\begin{align}
\kPPUD&=\bar{u}d(x) \bar{d}u(x)\bar{u}u(y)\bar{d}d(y),&&& \kUDPP&=\bar{u}u(x)\bar{d}d(x)\bar{u}d(y) \bar{d}u(y).
\end{align}
Every state of the local Hilbert space can be described by a set of $\Nf^2$ charges: baryon number $B\in\{-\Nf,\ldots \Nf\}$, isospin number $I\in \{-\Nc,\ldots \Nc\}$, diagonal flavor numbers $U,D,S,\ldots \in \{0,\ldots \Nc\}$ and for $\Nf>2$ additional generalizations of isospin $K_1, K_2, \ldots  \in \{-\Nc,\ldots \Nc\}$.
The one-link integral $J_k^{(B)}$ from Eq.~(\ref{JInt}) of order $k=n_1+2n_2+3n_3$ can be evaluated for various $\Nf$. We show here the general result for SU(3) in terms of $X=X_1$, $D_2$, $D_3$ and $E$: in black the $\Nf=1$ contributions, in blue the additional \blue{$\Nf=2$} contributions and in red the additional \red{$\Nf=3$} contributions:
{\scriptsize
\begin{align*}
J_0^{(0)}&=1,\qquad J_1^{(0)}=\frac{1}{3}  X,\qquad J_2^{(0)}=\frac{1}{12}  X^{2} - \blue{\frac{1}{24}  {D_2}},\qquad J_3^{(0)}=\frac{1}{36}  X^{3} - \blue{\frac{1}{24} X {D_2} } + \red{\frac{1}{60}  {D_3}},\nn
J_4^{(0)}&=\frac{1}{720}\lr{\blue{\frac{37}{12}  X^{4} - \frac{11}{2}   X^{2} {D_2}+ \frac{1}{6}  {D_2}^{2}} + \red{\frac{7}{3} X {D_3} }},\nn
J_5^{(0)}&=\frac{1}{840}\lr{\blue{\frac{47}{120}  X^{5} - \frac{31}{36}  X^{3} {D_2}+ \frac{1}{4} X {D_2}^{2} } + \red{\frac{1}{3} X^{2} {D_3}  - \frac{1}{9} \, D_{2} D_{3}}},\nn
J_6^{(0)}&=\frac{1}{448}\lr{\blue{\frac{53}{2700}  X^{6} - \frac{1}{18} X^{4} {D_2}  + \frac{5}{144} X^{2} {D_2}^{2}  + \frac{1}{45}  {D_3} X^{3} - \frac{1}{6480}  D_{2}^{3}} - \red{\frac{1}{40}  X D_{2} {D_3} + \frac{1}{240} {D_3}^{2}}},\nn
J_7^{(0)}&=\frac{1}{25920}\lr{\red{\frac{241}{2940}  X^{7} - \frac{19}{70}  D_{2} X^{5} + \frac{227}{1008}  {D_2}^{2} X^{3} + \frac{1}{9} \, {D_3} X^{4} - \frac{13}{1008}  D_{2}^{3} X - \frac{29}{168}  D_{2} D_{3} X^{2} + \frac{1}{168}  {D_2}^{2} D_{3}} + \red{\frac{11}{336} \, D_{3}^{2} X}},\nn
J_8^{(0)}&=\frac{1}{21772800}\left(\red{\frac{13259}{3360} \, X^{8} - \frac{149}{10}  D_{2} X^{6} + \frac{631}{40}  {D_2}^{2} X^{4} + \frac{121}{20} \, D_{3} X^{5} - \frac{403}{120} \, D_{2}^{3} X^{2} - \frac{143}{12}  D_{2} D_{3} X^{3}+\frac{1}{240} \, D_{2}^{4}}\right. \nn
&\left.\hspace{16mm}\red{ + \frac{11}{4} X {D_2}^{2} D_{3}  + \frac{11}{5} X^{2} D_{3}^{2}  - \frac{11}{20}  D_{2} D_{3}^{2}}\right),\nn
J_9^{(0)}&=\frac{1}{17107200}\left(\red{\frac{63163}{423360}  X^{9} - \frac{377}{588} X^{7} D_{2}  + \frac{1429}{1680} X^{5} {D_2}^{2}  + \frac{73}{280} X^{6} D_{3}  - \frac{1663}{5040} X^{3} {D_2}^{3}  - \frac{551}{840} X^{4} {D_2} {D_3}}\right.\nn
&\hspace{16mm} \left.\red{ + \frac{17}{3360}  X {D_2}^{4}  + \frac{71}{210} X^{2} {D_2}^{2} D_{3}  + \frac{13}{105} X^{3} {D_3}^{2}  - \frac{1}{420}  {D_2}^{3} {D_3} - \frac{13}{120} X {D_2} {D_3}^{2}  + \frac{13}{1260}  {D_3}^{3}}\right),\nn
\end{align*}
\begin{align}
J_0^{(1)}&=\frac{E}{6} ,\qquad J_1^{(1)}=\frac{E}{6}\lr{\blue{\frac{1}{4} X }},\qquad J_2^{(1)}=\frac{E}{6}\lr{\blue{\frac{1}{24} X^{2}  - \frac{1}{60}  D_{2}}},\qquad J_3^{(1)}=\frac{E}{6}\lr{\blue{\frac{1}{144} X^{3}  - \frac{1}{120} X D_{2}} + \red{\frac{1}{360}  {D_3}}},\nn
J_4^{(1)}&=\frac{E}{6}\red{\lr{\frac{1}{1344} X^{4} - \frac{1}{840} X^{2} D_{2}  + \frac{1}{20160}  {D_2}^{2}  + \frac{1}{2240} X {D_3} }},\nn
J_5^{(1)}&=\frac{E}{6}\red{\lr{\frac{143}{2419200}  X^{5} - \frac{29}{241920} \, D_{2} X^{3} + \frac{1}{32256}  D_{2}^{2} X + \frac{1}{23040} \, D_{3} X^{2} - \frac{1}{80640} D_{2} D_{3} }},\nn
J_6^{(1)}&=\frac{E}{6}\red{\lr{\frac{19}{4838400} X^{6} - \frac{221}{21772800} X^{4} {D_2}  + \frac{241}{43545600} X^{2} {D_2}^{2}  + \frac{11}{2903040} X^{3} D_{3}  - \frac{1}{21772800}  {D_2}^{3} - \frac{11}{3110400} X {D_2} {D_3}+ \frac{11}{21772800}  {D_3}^{2}}},\nn
J_0^{(2)}&=\blue{\frac{E^2}{144} },\qquad J_1^{(2)}=\blue{\frac{E^2}{144}}   \lr{\red{\frac{1}{5} X}},\qquad J_2^{(2)}=\blue{\frac{E^2}{144}}\lr{\red{\frac{1}{40} X^{2} - \frac{1}{120}  D_{2}}},\qquad 
J_3^{(2)}=\blue{\frac{E^2}{144}}\lr{\red{\frac{1}{360} X^{3} - \frac{1}{360} X D_{2} + \frac{1}{1260}  D_{3} }},\nn
J_0^{(3)}&=\red{\frac{1}{8640}  E^{3}}
\end{align}
}
\newcommand{\pu}{{\pi_U}}
\newcommand{\pd}{{\pi_D}}
\newcommand{\pp}{{\pi_+}}
\newcommand{\pq}{{\pi_-}}

The number of hadronic states in each conserved charge sector $(B,I,S,\ldots)$ can be obtained from the one-link integral by combining them
in alternating chains  $J_k^{(B)} J_{(N_f-B)\Nc-k}^{(B)}$, with $k$ denoting the mesonic occupation number $\meson_k$.
To obtain the Hamiltonian, we still have to integrate out the Grassmann variables. 
Here we consider the chiral limit only. The Grassmann constraint then dictates that all quarks $u,d,s,\ldots$ and anti-quarks $\bar{u},\bar{d},\bar{s},\ldots$ within mesons or baryons appear exactly $\Nc$ times. The Grassmann integral in the chiral limit for U(3) (i.e. SU(3) with $B=0$) for a given site $x$ is
\begin{align}
 I_G&=\int \prod_{\alpha}\prod_{f}[{\rm d}\bar{f}_\alpha {\rm d}f_\alpha ]\prod_{f,g} (\bar{f}g)^{k_{\bar{f}g}} 
(-1)^{\frac{1}{2}\sum\limits_{f\neq g}
k_{\bar{f}g}}F_\Nf(\Nc,\{k_{\bar{f}g}\})
\end{align}
with $k_{\bar{f}g}$ the sum of flavored dimers attached to site $x$, and 
with $F_\Nf(\Nc)$ some $\Nc$-dependent function, given the power $k_{fg}$ of charged fluxes ($f\neq g$). 
The result for non-zero baryon number $B\neq 0$ and details on $F_\Nf(\Nc)$ will be presented in a forthcoming publication. 
On a given configuration, the Grassmann integration simplifies due to flux conservation:
for the non-diagonal pseudo-scalar mesons $k_{\bar{f}g}=k_{\bar{g}f}$ such that all minus signs from the Grassmann integration cancel. 
Only the non-trivial contribution of type $k^{(2)}_{UD,\pi^+\pi^-}$ and its generalizations for $\Nf>2$ induce minus signs. Those link states can be resummed and diagonalized, e.g. the following states have an eigenvalue 1:
\begin{align}
\mPZ&=\lr{\kU\kD+\frac{1}{\sqrt{\Nc}}\kPPUD},& \mSZ&=\lr{\kP\kM+\frac{1}{\sqrt{\Nc}}\kUDPP} 
\end{align}
Likewise all dimer-based states are resummed, as states are only distinguishable on the quark level. This reduces the state space drastically to the physical Hilbert state, e.g. for $\Nf=2$: $B=0: 340\mapsto 50$, $B=\pm 1: 152\mapsto 20$, $B= \pm 2: 4\mapsto 1$, resulting in the local Hilbert space $\Hil$ as given in Tab.~\ref{HadronStatesNf2}, classified by baryon number 
$B$, isospin number $I$ and meson occupation number $\meson$.
The particle-hole symmetry generalizes to  $\meson  \mapsto\spin=\meson-\frac{\Nc}{2}(\Nf-|B|)$ as the meson raising and lowering operators fulfill an SU(2) algebra for each meson charge $Q_i$.
All states $\hadron\in \Hil$ for $\Nf=2,3$ contribute with a weight 1, although some dimer-based link weights are negative.
\begin{table*}
\begin{center}
{\footnotesize
\begin{tabular}{|r|r||c|c|c|c|c|c|c|c|c|c|c|c|c||c|}
\hline
$B$ & $I$\; & \multicolumn{13}{c||}{$\spin=\meson-\frac{3}{2}(2-|B|)$} & $\Sigma$ \\
\hline
 &  & $-3$ & $-\frac{5}{2}$ & $-2$ & $-\frac{3}{2}$ & $-1$ & $-\frac{1}{2}$ & $\, 0\,$ & $+\frac{1}{2}$ & $+1$ & $+\frac{3}{2}$ & $+2$ & $+\frac{5}{2}$ & $+3$ &  \\
\hline
\hline
 -2& 0 & & & & & & & 1 & & & & & & & 1 \\
\hline
 -1 & $-\frac{3}{2}$ &&&& 1 &&  1 && 1 && 1 &&&&  4 \\
 -1 & $-\frac{1}{2}$ &&&& 1 &&  2 && 2 && 1 &&&&  6 \\
 -1 & $+\frac{1}{2}$ &&&& 1 &&  2 && 2 && 1 &&&&  6 \\
 -1 & $+\frac{3}{2}$ &&&& 1 &&  1 && 1 && 1 &&&&  4 \\
\hline 
0 & -3 &  &&   &&& & 1 &&&&&&& 1\\ 
0 & -2 &  &&   && 1 && 2 && 1 && & && 4\\
0 & -1 &  && 1 && 2 && 4 && 2 &&  1 && & 10 \\
0 & 0  & 1&& 2 && 4 && 6 && 4 && 2 && 1 & 20 \\
0 & -1 &  && 1 && 2 && 4 && 2 &&  1 && & 10 \\
0 & -2 &  &&   && 1 && 2 && 1 && & && 4\\
0 & -3 &  &&   &&& & 1 &&&&&&& 1\\ 
\hline
 1 & $-\frac{3}{2}$ &&&& 1 &&  1 && 1 && 1 &&&&  4 \\
 1 & $-\frac{1}{2}$ &&&& 1 &&  2 && 2 && 1 &&&&  6 \\
 1 & $+\frac{1}{2}$ &&&& 1 &&  2 && 2 && 1 &&&&  6 \\
 1 & $+\frac{3}{2}$ &&&& 1 &&  1 && 1 && 1 &&&&  4 \\
\hline
 2 & 0 &&&&&&& 1 & & & & & & & 1 \\
\hline
\hline
$\Sigma$ &  &1 & 0 & 4 & 8 & 10 & 12 & 22 & 12 & 10 & 8 & 4 & 0 & 1 & 92\\    
\hline
\end{tabular}
}
\end{center}
\caption{
All 92 possible quantum states for the $\Nf=2$ Hamiltonian formulation with $\SU(3)$ gauge group. 
The number of states are given for the sectors specified baryon number $B$ and isospin number $I$, and symmetrized meson occupation number $\spin=\meson-\frac{\Nc}{2}(\Nf-|B|)$. Note the mesonic particle-hole symmetry  $\spin \leftrightarrow -\spin$ which corresponds to the shift symmetry by $\at$.
}
\label{HadronStatesNf2}
\end{table*}
Only single meson exchange is possible (multiple meson exchange becomes resolved into single mesons in the continuous time limit), 
the resulting interaction Hamiltonian has $\Nf^2$ terms:
\begin{align}
\hat{\calH}&=\frac{1}{2}\sum_{\langle\vec{x},\vec{y}\rangle}
 \sum_{Q_i} \lr{
 {\hat{J}_{Q_i,\vec{x}}}^+ {\hat{J}_{Q_i,\vec{y}}}^- + {\hat{J}_{Q_i,\vec{x}}}^- {\hat{J}_{Q_i,\vec{y}}}^+%
 }, 
\end{align}
e.g. for $\Nf=2$: $Q_i \in \{ \pi^+, \pi^-, \pi_U, \pi_D \}$ and for $\Nf=3$: $Q_i \in \{ \pi^+, \pi^-, K^+, K^-, K_0, \bar{K_0}, \pi_U, \pi_D, \pi_S \}$. 
For  the transition $\hadron_1 \mapsto \hadron_2$, the matrix elements $\expval{\hadron_1|Q_i|\hadron_2}$ of $\hat{J}_{Q_i}^{\pm}$ are determined from Grassmann integration and a square root per participating link weight (in, hopping, out); only those matrix elements are non-zero which are consistent with current conservation for all $Q_i$. Some matrix elements are negative, but in the combination of a closed charged meson loop the total weight remains positive.
Some examples for $\Nf=2$ are:
\begin{align}
\expval{\pi^+, \pi^-|\pi^+ |\pi_U ,\pi_D \pi^-}&=\frac{\sqrt{3}}{2},& \expval{\pi_U, \pi^-|\pi^+ |\pi_U,  2\pi^-}&=\frac{\sqrt{6}}{3},\nn
\expval{B_{uud}|\pi^+ |B_{uud},\pi_U ,\pi_D}&=\frac{2}{\sqrt{5}},& \expval{B_{uud}\pi^+ |\pi^+ |B_{uud},\pi_D}&=-\frac{\sqrt{3}}{6}
\end{align}
There are 
130 non-zero matrix elements for $B=0$ and 
40 matrix elements for $B=\pm 1$. The $B=\pm 2$ states do not allow for pion exchange.
Only after contraction of the matrix elements the resummation is carried out for the internal hadronic states, resulting in positive weights.
An important application of the $\Nf=2$ partition function is to determine the QCD phase diagram with both finite baryon and isospin chemical potential \cite{Nishida2003,Brandt2017}. 
Our formulation is still sign-problem free in the continuous time limit. As we have not yet performed dynamical simulations, we can only obtain analytic results in a high temperature expansion of the partition sum $Z$, which is an expansion around the static limit 
\begin{align}
Z\lr{\frac{\mu_B}{T},\frac{\mu_I}{T}}= & 2\cosh\frac{3\mu_{I}}{T}+8\cosh\frac{2\mu_{I}}{T}+20\cosh\frac{\mu_{I}}{T}+20\nn
&+ 2\cosh\frac{\mu_{B}}{T}\left( 8\cosh{\frac{\frac{3}{2}\mu_{I}}{T}}+12\cosh\frac{\frac{1}{2}\mu_{I}}{T}\right) +2\cosh\frac{2 \mu_{B}}{T}
\label{ZNf2Nc2}
\end{align}
in the number of spatial mesons. This will be presented in a forthcoming publication.

\definecolor{ar}{rgb}{0.0, 0.7, 0.0}

\section{Leading order gauge corrections to the Hamiltonian formulation for $N_f=1$}

\begin{figure}[h!]
\centering{
\includegraphics[width=0.6\textwidth]{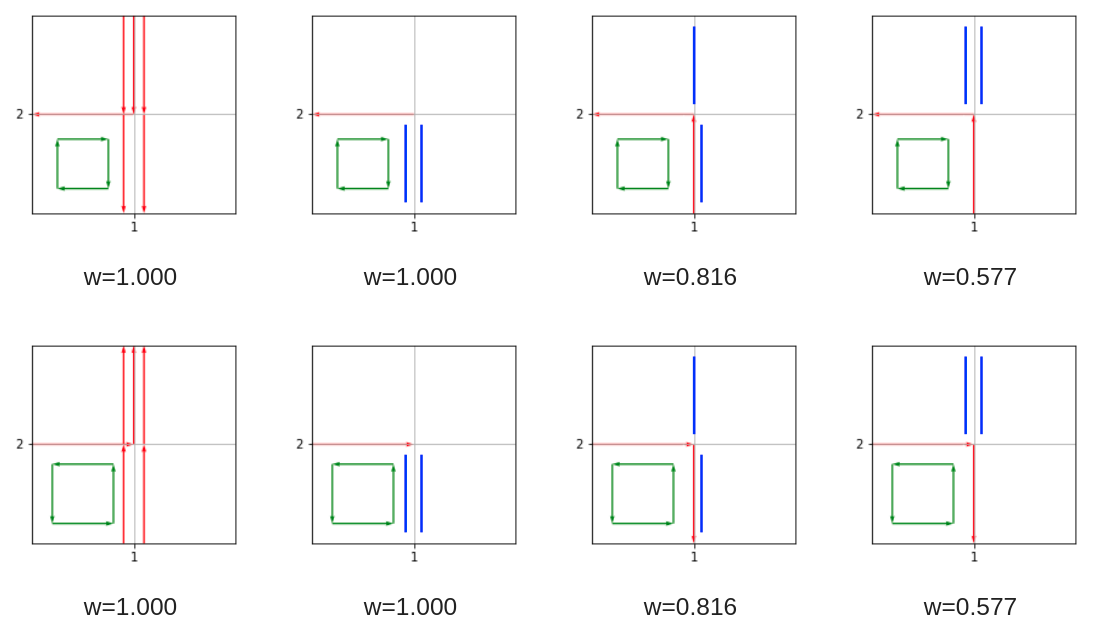}
}
\caption{Tensors and their weights of $\mathcal{O}(\beta)$ that can be incorporated into the Hamiltonian formulation.
Blue: dimers, red: baryon flux, green: plaquette flux due to gauge corrections.} 
\label{OB}
\end{figure}

The QCD Partition can be expanded via the strong coupling expansion in $\beta$:
\begin{align}Z_{QCD}&=\int d\psi d\bar{\psi} dUe^{S_G+S_F}=\int d\psi d\bar{\psi} Z_F \expval{e^{S_G}}_{Z_F},\nn
 \expval{e^{S_G}}_{Z_F}&\simeq 1+ \expval{S_G}_{Z_F} + \mathcal{O}(\beta^2)=1+\frac{\beta}{2\Nc} \sum_P \expval{{\rm tr}[U_P+U_P^\dagger]}_{Z_F}+ \mathcal{O}(\beta^2)
 \end{align}
Additional color singlet link states are due to plaquette excitations. For discrete time lattices, the dual formulation has been extended beyond $\mathcal{O}(\beta)$ in terms of a tensor network \cite{Gagliardi2019}. Here we want to consider the leading order gauge corrections 
in a Hamiltonian formulation.  On anisotropic lattices, the anisotropy $\xi=\frac{a_s}{a_t}$ is a function of two bare anisotropies 
$\gamma_F$ and $\gamma_G=\sqrt{\frac{\beta_t}{\beta_s}}$, i.e $\xi=\xi(\gamma_F,\gamma_G,\beta)$.
 However, in the continuous time limit $a_t\rightarrow 0$ ($\xi \rightarrow \infty$) and for small $\beta$, spatial plaquettes are suppressed over temporal plaquettes by $ (\gamma_G\gamma_F)^{-2}$, hence only temporal plaquettes need to be considered. They are of the same order as meson exchange, hence the $\mathcal{O}(\beta)$ weights will contribute to the corresponding operators $\hat{J}^{\pm}$. Note that $\hat{J}^{\pm}$ still has block-diagonal structure, but they will also allows to couple to baryons.
 Fig.~\ref{OB} shows the relevant weights as computed in \cite{Gagliardi2019}.

\section{Summary and Outlook}

We have presented two extensions to the Hamiltonian formulation of lattice QCD: (1) generalization from $\Nf=1$ to $\Nf>1$ in the strong coupling limit, (2) $\mathcal{O}(\beta)$ gauge corrections for $\Nf=1$.
In a forthcoming publication, we also address gauge corrections to the $\Nf=2$ Hamiltonian. It turns out that - although not sign problem-free - the sign problem is much milder than in the corresponding formulation at finite $\Nt$ as for $\Nt\rightarrow \infty$ only temporal plaquettes contribute. 
Meson exchange between baryons can thus be either due to the gauge corrections or the flavor content.
All results presented here are valid in the chiral limit. We have argued in \cite{Klegrewe2020} that also a Hamiltonian formulation at finite quark mass $am_q$ is well-defined, which extends this approach to staggered lattice QCD further. We are preparing Quantum Monte Carlo simulations that will help to obtain the QCD phase diagram on the lattice in the parameter space 
$(aT,a\mu_B,a\mu_I,am_q,\beta)$ in lattice units or in units of the baryon mass $am_B$.

This work was supported by the Deutsche Forschungsgemeinschaft (DFG, German Research Foundation) – project number 315477589 – TRR 211.“

\end{document}